\documentclass[12pt]{mod_elsarticle}

\usepackage{amssymb}
\usepackage[latin1]{inputenc} 
\usepackage[T1]{fontenc}      
\usepackage{geometry}         
\usepackage[english]{babel}  
\usepackage{graphicx}
\usepackage{vmargin}
\newtheorem{thm}{Theorem}
\newtheorem{lem}[thm]{Lemma}
\newdefinition{rmk}{Remark}
\newproof{pf}{Proof}
\newproof{pot}{Proof of Theorem}

\journal{.}

\begin{document}

\begin{frontmatter}



\title{Emergence of cooperation, selection of interactions and network formation}


\author{David Chavalarias}
\ead{david.chavalarias@polytechnique.edu}
\ead[url]{http://chavalarias.com}
\address{Institut des Systèmes Complexes de Paris Ile-de-France\\
57-59 rue Lhomond F-75005 Paris, France}

\begin{abstract}
We propose a model of \textit{emergence of cooperation} in evolutionary games that highlights the role of \textit{network formation} and effect of \textit{network structure}. In line with empirical data, the model proposes a mechanism that explains the \textit{persistence of heterogeneous types} (heterogeneity in rules for changing behavior) within a population, and in particular the \textit{sustainability of altruism} (presence of unconditional cooperators) even in case of \textit{strong social dilemma}. This explanation constitutes an alternative to the choice and refusal mechanism that is often presented as an explanation for cooperation on dynamic networks. We also exhibit a minimal set of strategies for emergence of cooperation (altruists, reciprocators and selfish agents) and sketch a two steps scenario for this emergence.  We adopt an \textit{hybrid methodology } with both analytical and computational insights. 	

\end{abstract}


\end{frontmatter}

\section*{Introduction}
Emergence of cooperation under natural selection pressure as long been a puzzle for evolutionary game theory, which traditionally considers agents interacting at random within the population. However, new insights have been obtained since modelers  recently started to
take into account networks and their dynamics in the modeling of interactions.
Besides structural and topological constraints that determines the possibilities and conditions of interactions (who can interact with who), \emph{activity networks} (who is actually
interacting with who) are emergent patterns of social activities. People meet
each other, interact, learn from these interactions and eventually keep previous partners in mind as potential future partners.
These  emergent networks from agent's activities strongly influence their outcomes, thus leading to  a reciprocal dependency between agents' activities and the networks they form.

To give some insights into the interplay between social activities and the underlying evolving social networks, this papers analyzes a model of emergence of cooperation on dynamic networks. We choose the prisoner's dilemma game as a case study since it is known to be the worst situation for cooperative dyadic outcomes, compared to other related games such as the Chicken game.

We determine the conditions for emergence of cooperation from an all defecting population and show how heterogeneous population states, with both cooperators and defectors, can be stable thanks to emergent structures of the activity network. This work builds on a model previously introduced \cite{ahn:endo} and proposes an alternative to the cliquishness or  "choice and refusal mechanism" that often stand for explanations of cooperation.

\section{The prisoner's dilemma on networks}

 A prisoner's dilemma game (PD game) is a two players game which payoffs matrix has a
 structure described by table 1 with  the conditions $0<p<r<1$.

\begin{table}
  \centering
  \caption{The matrix of the prisoner's dilemma game}\label{Table 1}
\begin{tabular}{|c|c|c|} \hline
Player $A$ & \multicolumn{2}{c|}{Player $B$}\\
& $C$ & $D$ \\ \hline $C$ & $(r,r)$ & $(0,1)$ \\ \hline $D$ &
$(1,0)$ & $(p,p)$\\ \hline
\end{tabular}

\end{table}

The prisoner's dilemma game belongs to the class of social dilemmas and has become a paradigm for modeling cooperation and altruistic behavior, \textit{i.e.} situations where one achieves an action that benefits to the other at its own expense. The dilemma lays in the fact that option $D$ (defect) leads always to the highest reward whatever the other does. But when both players choose $D$, they finally receive less than if both had played $C$ (since $p>r$). This means that mutual cooperation is always more profitable than mutual defection (collective rationality), but given the opponent's action, defection is always individually more profitable than cooperation (individual rationality). The prisoner's dilemma is thus a prototype of situations where there is a dilemma between individual and collective rationalities.

The game can be played several times in a row or only once ; players can play simultaneously or sequentially. In this paper, we will study repeated sequential PD game. The repeated sequential form of the prisoner's dilemma can be useful to model social interactions based on the mutual gift, as described for example in \cite{Mauss2001Gift}, or trust (\cite{dasg:trus}, \cite{haya:reci}). The recent access to huge databases related to sequential social interactions (phone calls, mails, posts on blogs, etc.) constitute also a new source of applications of PD game modeling (\cite{Lozano2008Mesoscopic}), with perspectives of empirical validation.

Experimental studies based on the prisoner's dilemma game generally conclude that the
level of cooperation observed in  human population is not consistent with non-cooperative
game theory predictions (\cite{clar:sequ}; \cite{haya:reci}
for the sequential form). In particular, one of the
most puzzling phenomena observed in experimental studies is that
among the various behaviors encountered, there is often a substantial
proportion of altruist behaviors, \emph{i.e.} \textit{unconditional cooperators}
\cite{ahn:coop}. Behaviors observed in experimental studies can be roughly
categorized into three types: selfish players, who
most of the time defect, conditional cooperators, who cooperate if
the other player did, and altruist players, which most of the time
cooperate whatever the other did. If heterogeneity in agent's types is addressed in some models, few of them explain how it can be maintained, and even fewer explain how it can emerge from an homogeneous population of selfish players.

Recent literature demonstrated that the topology of social networks
can help to sustain cooperation (\cite{bata:evol}
; \cite{ashl:pref} ; \cite{hrus:frie}; \cite{Eguiluz2005Cooperation}). One of the
main conclusions of these studies is that cooperation is favored by
a high degree of cliquishness of the activity networks and the possibility of refusing interaction.  This is a
motivation to consider models with partner selection where the topology  and weights of the activity network are emergent properties.

In the following, we will propose a mechanism of endogenous network formation that could stand for an alternative to the cliquishness or "choice and refusal" mechanism in the explanation of emergence and sustainability of cooperation with heterogeneous population.

\subsection{The model: why address books are important}
Let $P$ be a population of $N$ agents playing PD game in discrete-time. The game is played sequentially: an agent (the first mover) announce its play ($C$ or $D$), and its partner (the second mover) reacts accordingly. For example, someone can post a message on the blog of a friends, and this latter can answer (or not) with a post on the first mover's blog. We will assume that at each period, each agent has the opportunity to play only once as a first mover (this can be interpreted as a time constraint).

Agents keep in their address book the names of their partners for all previous successful interactions as first mover ($(C,C)$ or
$(D,C)$ outcomes). Names in agent $i$ address book will be called its \textit{ relationships } and agents having $i$ in their address book will be called its \textit{ friends}.

We will consider a  complete graph for the network of accessibility (first movers can choose any agent in the population, like for mail, comments post or phone calls). The network of activity defined by the set of all address books will be an emergent evolving directed weighted network.

Agents can choose their partners to some extent:
 \begin{itemize}
   \item With a probability $(1-e), e\in[0,1]$, first movers can choose their partners in their address book, thus taking advantage of what they have learned so far. If their address book if empty, they are matched at random in the population.
  \item  With a probability $e$, first movers are matched at random in the population.
 \end{itemize}
Parameter $e$ can be interpreted as an uncertainty level in the partner selection process or as an exogenous structural factor (\textit{e.g.} some competencies an agent needs are sometime not present in its address book). We will notice that $e=1$ corresponds to the classical approach of evolutionary game theory for which is as been proved that cooperation is not sustainable.

Agents start their life with empty address books and do friends and relationships during social interactions.

\subsection{The strategies}
A period of the game unfolds as follows: \begin{enumerate}
  \item  Each agent is matched as first mover with a partner according to rules given above,
  \item  Each agent does a proposition ($C$ or $D$) to its partner,
  \item Each agent answers ($D$ or $C$) as a second mover to every agent that has
chosen it.
\end{enumerate}
At a given period, an agent can thus play several times as a second mover, but only once as first mover.

 The set $S$ of strategies is constituted of the three types mentioned previously: altruist,
reciprocator and selfish agents in proportion $\gamma$, $\delta$ and $1- \gamma  -\delta$. These types are defined by their behaviors as first mover (table 2) and as second mover (table 3). Their proportions evolve in the simplex $\triangle = \{ (\gamma,\delta)|\delta+\gamma \leq 1 ; \gamma \geq 1 ; \delta \geq 1 \}$ .  We will assume that agents know the proportions of the different types in the population\footnote{More sophisticated model would assume that agents learn these proportions from the outcome of their previous interactions} so that they can compute their expected payoffs $E$ as first mover and can play accordingly : $E(C)=(\gamma+\delta)r$ ;
$E(D)=(1-\gamma)p +(\gamma)$. The condition under which
$E(C)>E(D)$ is thus $\delta>\frac{p}{r}+\frac{(1-p-r)\gamma}{r}$

Although they are stylized types, these strategies will give some insight into the phenomena that could potentially stabilize cooperation.

\begin{figure}
  \includegraphics[width=200pt]{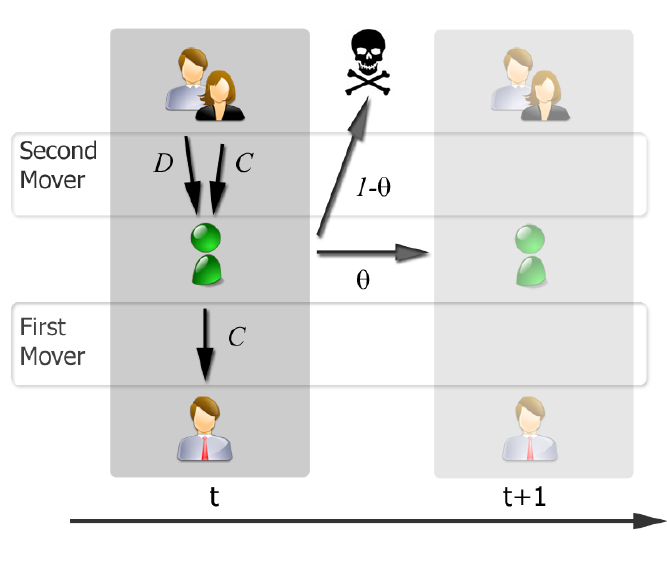}\\
  \centering
  \caption{Interactions as first and second movers. Each round, an agent can play several times as a second mover, but only once as first mover.}\label{epspace}
\end{figure}

\begin{table}\footnotesize
  \centering
  \caption{Behaviors of first movers in function of the state $(\gamma,\delta)$ of the population.}\label{Table 3}
\begin{tabular}{|c|c|c|} \hline
 & \multicolumn{2}{c|}{First Mover}\\
Type & In address Book & random  \\
& (with relationships) & \\
\hline Selfish & Search for an altruist and play $D$.  & $C$ if $\delta>\frac{p}{r}+\frac{(1-p-r)\gamma}{r}$ \\
& If none play $C$&  \\
\hline Reciprocator & Pick a name at random  & $C$ if $\delta>\frac{p}{r}+\frac{(1-p-r)\gamma}{r}$ \\
 & and play $C$ & \\
\hline Altruist & Pick a name at random   & $C$ \\
 & and play $C$ & \\
\hline
\end{tabular}

\end{table}

\begin{table}\footnotesize
\centering \caption{Behaviors of agents as second
movers.}\label{Table 2}
\begin{tabular}{|c|c|c|c|} \hline
 Type &  Second Mover \\
\hline Selfish & $D$ \\
\hline Reciprocator & Do what first mover did.\\
\hline Altruist & $C$ \\
\hline
\end{tabular}

\end{table}

\subsection{The evolutionary dynamics}

At the end of a period, fitness of agents are defined as the
sum of all their dyadic  interactions payoffs at this period.
These payoffs are assumed to be linked to a reproductive success: agents
have a probability $(1-\theta)$ to die at the end of
 period.

 Each died agent is replaced by a new agent according to a replicator dynamics so that the population size is kept constant: with a probability proportional to its fitness, one agent is chosen at random in the population, and a new agent of the same type  with an empty address book is created (\emph{overlapping generation game}).

\section{Analytical and computational study}
We will now give some  analytical and computational insights into this model. One of the stylized fact we want to reconstruct is the existence of areas in the $(e,p)$ parameter space where we find stable heterogeneous population with significant level of cooperation. This means that the dynamics under study shall have an internal attractor in $\Delta$.

Given the complexity of the model, an analytical study will necessarily adopt some approximations ( \textit{e.g.} infinite population size assumption). On the other hand, computational studies can never guaranty that the number of time-steps considered in simulations is sufficiently high to distinguish a very slow evolution toward an homogenous population from the existence of an internal attractor. When the vector fields tends to zeros and when there no way to assess convergence, simulations are poorly informative about internal attractors.

Our methodology is thus to find an analytical approximation of the model (2.1) and check that it fits well the simulated model (2.2). Then understand cooperation in the analytical model (3.1), characterize internal attractors (3.2) and study one of these internal attractors to understand how network patterns can sustain cooperation in heterogeneous population (3.3). Then we will see to what extent the choice of the strategy set is important to observe these phenomena (4).

\subsection{The fitness function}
We can see from table 2 that behaviors of agents vary in function of the sign of  $\delta-\frac{p}{r}$. The part of $\Delta$ concerned by
the problem of emergence of cooperation contains the point $(0,0)$ corresponding to an homogeneous
populations of selfish agents. The interesting area for our study is
thus defined by $\delta<\frac{p}{r}+\frac{(1-p-r)\gamma}{r}$. For sake of clarity, we give in
this paper the analytical approximation of expected payoffs for this area only.

\begin{thm}{\label{theo1}}
 \emph{For an infinite
population defined by $(\gamma,\delta)\in\Delta$, with the approximation that the proportions of types
in population are constant at the agents' time
scales, the mean expected payoffs $\pi_{a}$, $\pi_{r}$ and $\pi_{s}$  (for altruists, reciprocators and selfish agents respectively) in the domain $\delta<\frac{p}{r}+\frac{(1-p-r)\gamma}{r}$ are given by:}

   $$ \pi_{s} = (1-e)(\frac{2p(1-\gamma+\theta(\gamma-1))+\gamma}{1+\theta(\gamma-1)}+
    \frac{(1-\theta)\gamma}{1-(1-(\delta+\gamma))\theta})+2e(p(1-\gamma)+\gamma) \\
$$

$$  \pi_{r} =(1-e)(\frac{\theta\gamma(r+2p-1)+\gamma+2p(1-\gamma-\theta)}{1+\theta(\gamma-1)}+
  \frac{r\gamma}{1-(1-(\delta+\gamma))\theta})+e(2p+\gamma(1-2p+r))\\
$$

$$      \pi_{a} = (1-e)(\frac{r\theta\delta}{1+\theta(\gamma-1)}+
  \frac{r(2\gamma+\delta)}{1-(1-(\delta+\gamma))\theta})+er(\delta+2\gamma)\\
$$
\begin{pf}{\ref{theo1}} Proof given in appendix. \end{pf}.

\end{thm}

From this payoffs expressions, the relative
fitness for each type is given by:
\begin{equation}\label{fitness}
    f_i=\frac{\pi_i-<\pi>}{<\pi>}
\end{equation}

where $<\pi>=\gamma \pi_a + \delta \pi_r + (1-\delta-\gamma)\pi_s $

\bigskip

Whether cooperation will develop in the population will depend on $e$, $r$ and $p$. For sake of convenience and clarity we will adopt a standard parametrization of the PD game matrix:
$r=(1-p)$. This enables to reduce the parameter space to two
free parameters $e$ and $p$ which helps for visualization. Intuitively, $p$ fixes the strength of the social dilemma: the higher $p$, the stronger the dilemma. In the following, attractors and analytical trajectories are computed by solving these equations numerically.

\subsection{Comparison between analytical and computational results}

We  first can compare the dynamics of the analytical model and the simulated model. We present here the comparison for the parameters: $\theta=0.99$, $e=0.3$  and $p=0.3$ (thereafter referred as settings 1). We ran 10 simulations and collected data on types' proportions in population, address books and payoffs.  Initial conditions are
$(\gamma,\delta)=(0.1,0.1)$ and $N=500$.  Initial rate of cooperation for first move play is thus $10\%$.

\begin{figure}
  \includegraphics[width=300pt]{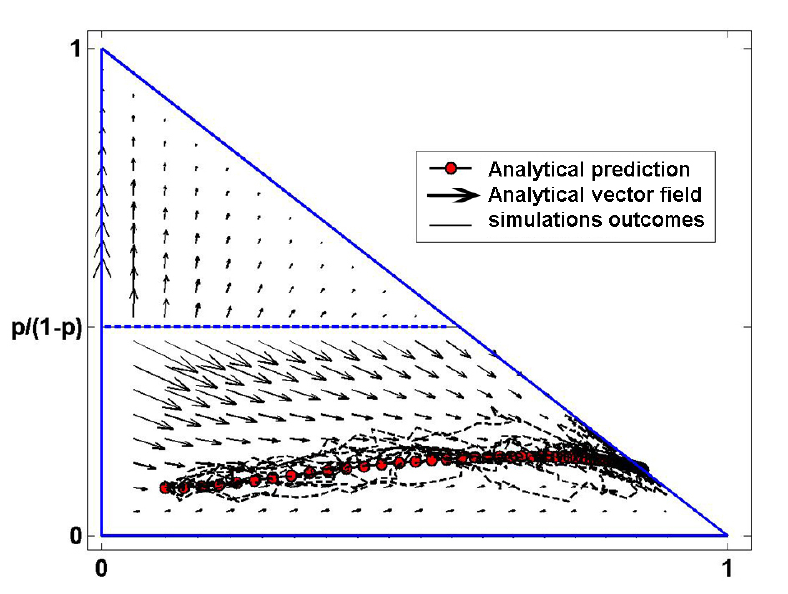}
  \centering
  \caption{Comparison between analytical predictions and simulations. For $e=0.3$,
   $p=0.3$, $\theta=0.99$ the theoretical analysis predicts
  very well the behavior of the system. \emph{Line with circles:} Analytical prediction of the trajectory  in case of infinite population size. \emph{Dotted lines:} plots of $10$ independent
   simulations with $N=500$. Each circle is separated by $100$ periods i.e. the agents'
     mean lifetime expectancy.}\label{trajectoire}
\end{figure}

As expected from the analytical study (fig. \ref{trajectoire} - circles on the graph), the population
converges toward a full cooperative state with a mixed population of altruists and reciprocators. Cooperation has become sustainable and selfish agents have disappeared.

We can also check that the stationary environment approximation
made in proposition 1 is quite reasonable: the
time between two markers (circles) represents one generation. The
proximity of two successive markers indicates that the environment
of an agent does not change much during its lifespan. This approximation will always be satisfied around an internal attractor since it is the place where the vector field tends to zero (the fitness function is continuous in $\Delta^\circ$).

\section{Cooperation,  social activity and network formation}

We will now see what we can learn from this model and its infinite population size approximation about the role of network formation in emergence of cooperation.

\subsection{Cooperation and social activity}
To better understand the reasons of the convergence towards a fully cooperative state in settings 1, we can turn toward the analysis of the structure of social interactions for each type of agent in the infinite population approximation. From the analytical study, we can compute the theoretical level of activity for each type in function of $\delta$ and $\gamma$, \emph{i.e.} the mean number  of interactions per period. This provides information about the degree of involvement in social interactions for each type of agent. Let's define the relative activity for altruists $RA_a$ (resp. reciprocators $RA_r$) as the ratio between
the mean number of interactions per period for altruists (resp. reciprocators) and the mean number of interactions per period for selfish agents as a function of $(\gamma,\delta)$.

Details about the computation of these relative activities are given in appendix. We can observe on figure 3 that: \emph{a)} these two ratios are always greater than one, which means that
selfish agents are the less active in the social network (they don't
have friends actually); \emph{b)} altruists are the most active of all types, especially when
in small proportion.

\begin{figure}
  \includegraphics[width=200pt]{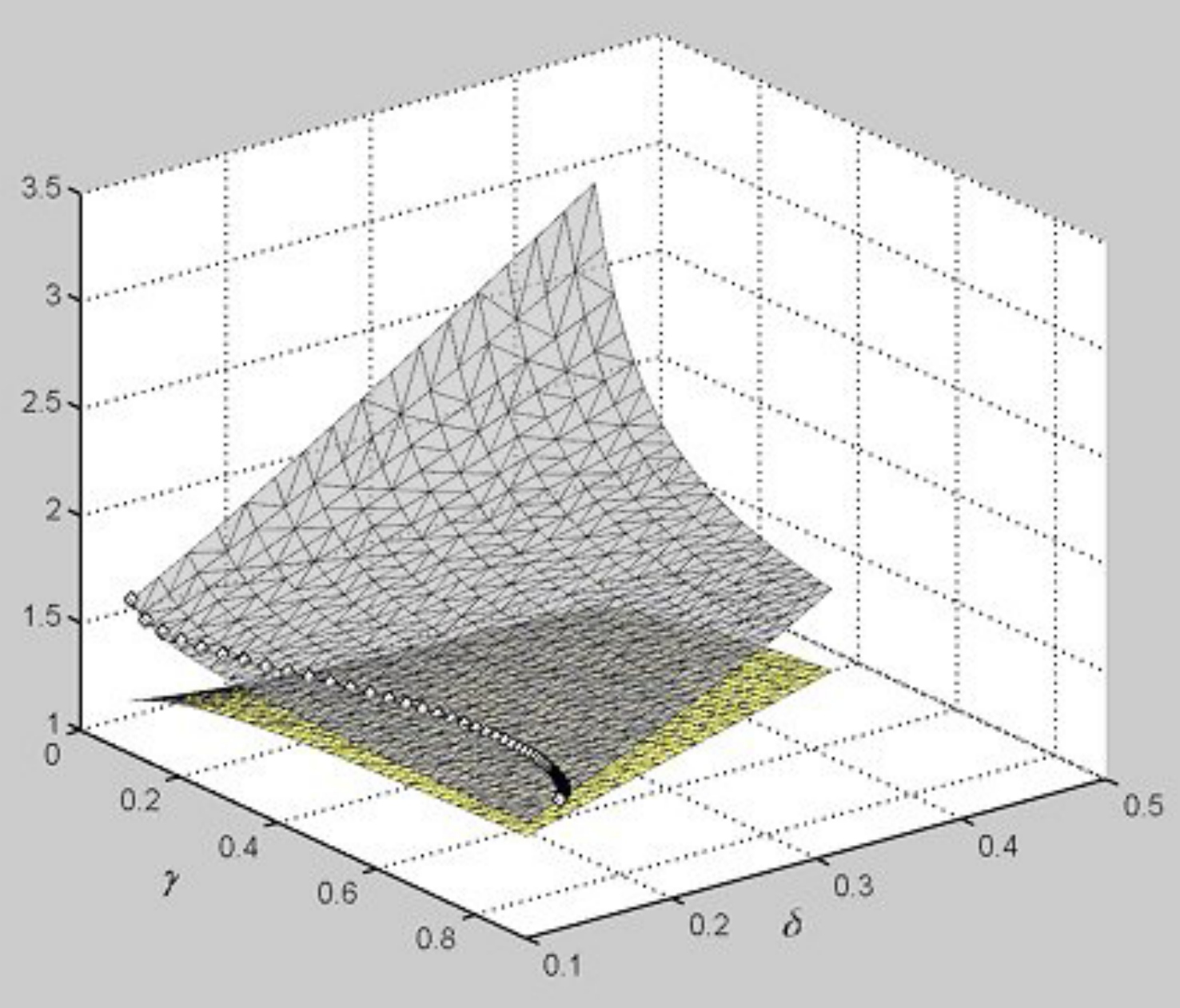}\\
  \centering
  \caption{Theoretical relative activities for reciprocators and
altruists ($0.1<\gamma<0.9$ and $0.1<\delta<0.4$ ; $e=0.3$
  and $p=0.3$, $\theta=0.99$). \emph{Upper surface :} relative activity of
  altruists ; \emph{lower surface:} relative activity for reciprocators}\label{activity}. The plotted trajectory is the  projection on this surface of the theoretical trajectory of figure \ref{activity}
\end{figure}

This example illustrates the way altruists can be successful even in presence
of selfish agents: they are more active than everybody else. Since
altruists are the only ones to cooperate as second mover, they are the only one to become friends with
reciprocators, and consequently they do relationships faster than other types. Altruists
are more popular. The fact that their selfish partners cheat them is
balanced by the fact that they are more appreciated as second mover
partners in the social network.

\subsection{Looking for heterogeneous equilibriums}

An interesting question concerning the infinite population size approximation concerns the existence of internal attractors. What about the stylized fact presented in the introduction: stable heterogeneous populations of altruists, reciprocators and selfish agents? Given
$(e,p,\theta)\in ]0,1[\times]0, 0.5[\times]0,1[$, the replicators
dynamics is determined by the folowing vector field $V$ in the simplex
$\Delta$:
$$    V_\delta =\frac{\delta\pi_r}{\gamma\pi_a-\delta\pi_r-(1-\delta-\gamma)\pi_e}-\delta\mid_{e,p}
$$

$$
    V_\gamma =\frac{\gamma\pi_a}{\gamma\pi_a-\delta\pi_r-(1-\delta-\gamma)\pi_e}-\gamma\mid_{e,p}
$$

A point
$(\delta,\gamma)\in\Delta^\circ\cap(]0,1[\times]0,\frac{p}{(1-p)}[)$
is an internal attractors only if:
\begin{equation}\label{internal}
    V_\gamma^2+V_\delta^2=0\mid_{e,p}
\end{equation}

We solved numerically equation \ref{internal} to get the set of internal
attractors for a given value of $\theta$. Figure 4 displays the set
of pairs $(e,p)$ for which the dynamics has an internal
attractor. This set has a non-empty interior and for $\theta=0.99$ it
is crab's claw shaped and oriented along the
line $p=0.5-0.35 e$. As shown on the figure, the rate of cooperation at the
attractor decreases in the population along this line, which means
that, as expected, $e$ has a negative impact on cooperation. Figure 5 displays the
set of pairs $(\gamma,\delta)$ that are internal attractors for
some value of $e$ and $p$. We can see that most of these attractors
are situated above the line $y=x$, which means that there are generally
more reciprocators than altruists at internal attractors. This is in line with experimental studies.

Last, we can  study the influence of the life expectancy, which is also the ratio between the network updating timescale and the population renewal timescale. If cooperation disappears for very low life expectancy (\textit{e.g.} $\theta=0.5$), we observe cooperation  and internal attractors as soon as $\theta=0.9$, which means that cooperation is quite robust against variations of life expectancy.

\begin{figure} \includegraphics[width=200pt]{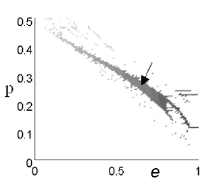}\\ \centering \caption{Existence of internal attractors in the $(e,p)$ space. The level of gray is indexed on the rate of cooperation at the attractor (the darker, the less cooperative). This has been obtained from Equ. \ref{internal} solved by numerical analysis for $\theta=0.99$. The arrow points to the pairs $(e,p)$ corresponding to the simulation presented in fig. 6.}\label{epspace} \end{figure}

\begin{figure} \includegraphics[width=200pt]{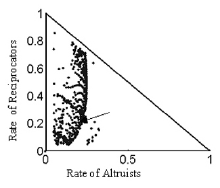}\\   \centering   \caption{Location of attractors in the $(\gamma,\delta)$ space obtained from numerical solution of Equ. \ref{internal}. Each point represents an attractor for some value of $(e,p)$ (fig 4.).}\label{attractor} \end{figure}

\subsection{Patterns in the emergent network}

With this map of attractors from the analytical study, we can look at the structure of internal attractors in the simulated model and check, passing, that the analytical model predicts well the existence and location of internal attractors for the simulated model. For example, a simulation ran for 25.000 time steps for
$\theta = 0.99$, $e=0.7$ and $p=0.25$ (thereafter referred as settings 2 - the arrow
on fig. 5) has a trajectory that cycles around the analytically predicted attractor, which corroborate the existence of an internal attractor despite the hight value of $e$. Its location is close to the one predicted by this analytical insight (fig. 6).

\begin{figure}  \includegraphics[width=300pt]{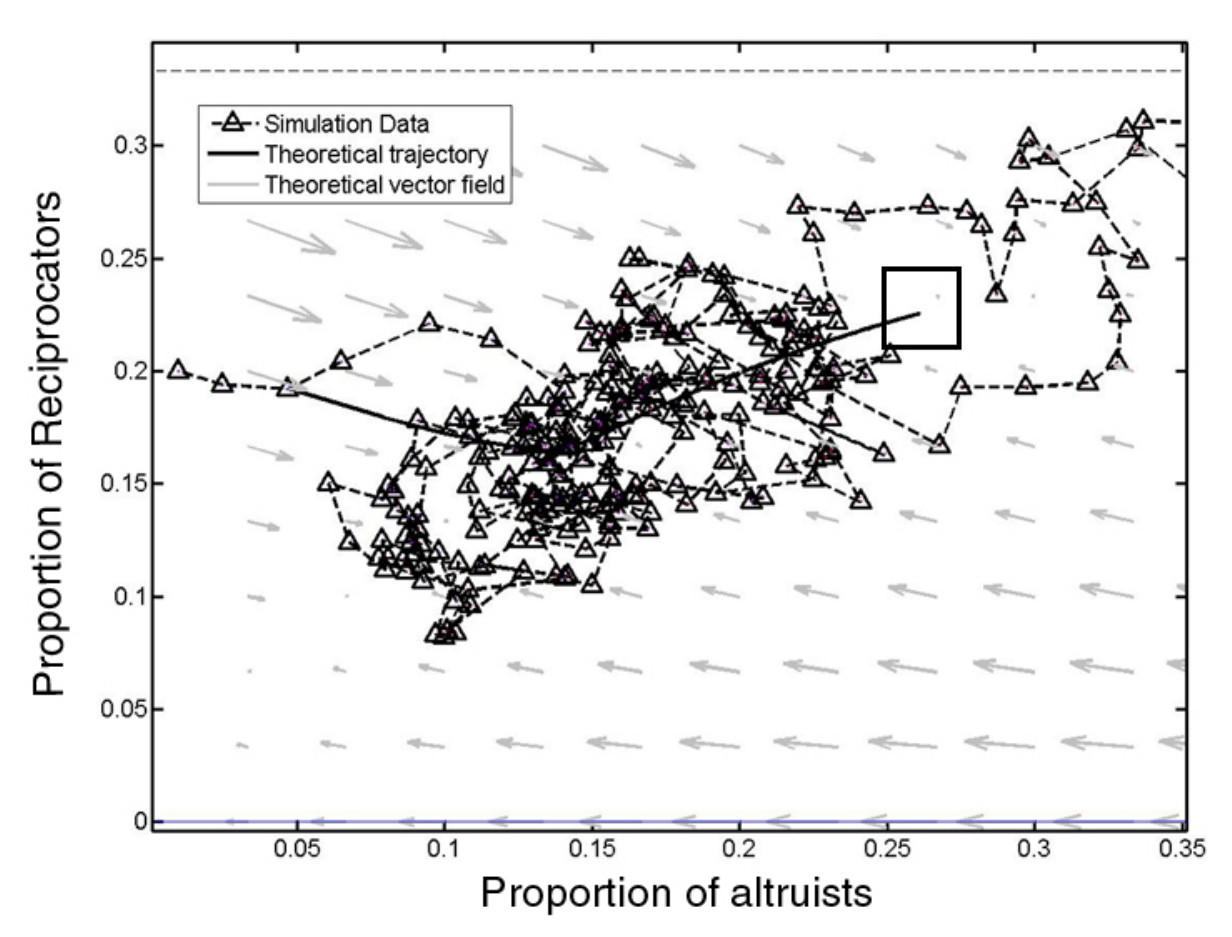}\\
  \centering
  \caption{A case of internal attractor ($e=0.7, p=0.25, \theta=0.99$) : For some values of $e$ and $p$, a society starting from an almost all-defecting state ($19\%$ of D-first-mover reciprocators, $80\%$  of selfish agents, $1\%$  of altruists ) evolves toward an internal attractor with a mixed population of the three types. The simulation presented here is 25 000 periods long, corresponding to 250 mean lifetime expectancy. The black line represents the theoretical evolution in an infinite population and the theoretical attractor is situated in the center of the black square. With finite population size (here 1000 agents), the population cycles around the attractor. Asymmetries in the vector field's strength around the theoretical attractor as well as variance effects might explain the discrepancy between simulation results and predictions.}\label{internal attractor}
\end{figure}

The analysis of the structure of the emergent social network at this internal attractor helps us
to understand the interplay between the dynamics of the activity network and emergence of cooperation. In directed networks, important vertices can be defined in terms of hubs and authorities\footnote{If a vertex points to many vertices with large authority weight, then it should receive a large hub weight. If many vertices with large hub weight point to a vertex, this latter should receive a large authority weight.} \cite{klei:hubs}.

In our model, authorities are agents with lots of friends, and hubs are agents with lots of hub relationships. The plot of the directed network formed by the address books around the attractor (settings 2) reveals an interesting structure: the first hubs are always of altruist type, the first authorities are always of reciprocator or egoist type. For example, figure 7 represents the ten hubs and ten authorities of a network of 100 vertices with $\theta=0.99$, $e=0.7$ and $p=0.25$. All the ten authorities are altruist while the ten hubs are all of selfish or reciprocator type.

This kind of organization of the social network suggests an explanation of cooperation than differs from the cliquishness of cooperators: \emph{popularity of cooperators}. The details of the
activity equations in the infinite population approximation support this view. We can see that while altruists are more likely to be authorities, selfish agents and reciprocators are more likely
to be hubs.  This emphasizes the importance of having the three
types of agents to get heterogeneous equilibriums at behavioral level.
Types maintain for different reasons: altruists because they have a  lot of friends and among them both altruists and reciprocators; reciprocators because they avoid being exploited by selfish agents and still have some friends among altruists, selfish agents
because they exploit altruists.

We will notice passing, that this simulation is also a case of emergence of cooperation from an initial  population that had only $1\%$ of altruist agents. In most models explaining cooperation, especially models based on cliquishness, sustainability only is explained because the model requires that the proportion of altruists passes a certain threshold to be sustainable.  In these cases, emergence cannot be explained.

\begin{figure}
  \includegraphics[width=300pt]{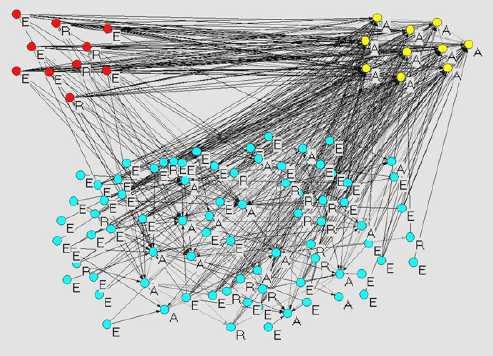}   \centering
  \caption{An example of network structure in the case of an internal attractor (100 agents, $\theta=0.99$, $e=0.7$ and $p=0.25$). Arrows indicate who is the friend of whom. The graph shows the ten hubs and ten authorities of the network (a Pajek plot). Authorities are all of altruist type ($A$) while hubs are all of selfish ($E$) or reciprocator ($R$) type.}\label{hubs} \end{figure}

\section{Scenarios for emergence of cooperation}
Our initial aim was to explain the sustainability of altruists in presence of selfish agents. We can wonder whether the presence of  reciprocators is contingent or necessary. It might be hard to study this question from a  computational point of view because, as a matter of emergence, a simulation can only prove that emergence is possible (by the presentation of a simulation where emergence took place) but not that it is impossible. We will thus turn again toward the analytical model to get some insight.

\begin{thm}{\label{theo2}} For an infinite population of selfish and altruist agents, an all-selfish population is an evolutionary stable state (ESS).

\begin{pf}{\ref{theo2}}
If we develop $\pi_{altruists}-\pi_{selfish}|_{\delta=0}$ in power of $\gamma$ at the origin we have:

\begin{equation}\label{origin}
    \pi_{altruists}-\pi_{selfish}=-(1-e)\frac{2p(1-\theta)}{1-\theta}-2ep+(1-e)
    \frac{\theta}{1-\theta}\gamma+\circ(\gamma)
\end{equation}

The first coefficient of this development is always negative and the second always positive. This implies that a population of selfish agents is an evolutionary stable state (ESS) when selfish and altruist are the only available strategies.
\end{pf}.

\end{thm}

The consequence of theorem \ref{theo2} is that emergence of cooperation is never possible through introduction of few altruists in large population of selfish agents (except in the trivial case $p=0$). Nevertheless it often can maintain for some values of $e$ and $p$ once this proportion passes a certain threshold.

\begin{thm}{\label{theo3}} For an infinite population of selfish agents and reciprocators, the evolution is neutral between the two types in the area $\delta<\frac{p}{r}+\frac{(1-p-r)\gamma}{r}$.

 \begin{pf}{\ref{theo3}} The proof of this theorem is straight forward since for $\delta<\frac{p}{r}+\frac{(1-p-r)\gamma}{r}$, reciprocators and selfish agents have the same rule of behavior and thus $ \pi_{r}=\pi_{s}$.

\end{pf}.

\end{thm}

The study of the fitness expression in the domain $\delta>\frac{p}{1-p}$) shows that above this threshold, reciprocators are sustainable even if they are first mover cooperators. Since this threshold is quite high, and given that neutral evolution is a very slow process, it is unlikely that cooperation emerges that way in large populations.

The consequences of theorems \ref{theo2} and \ref{theo3} is that the most probable scenario for emergence of
cooperation starting from an all-selfish population is two folds: 1) a slow neutral evolution from a mixed population of selfish agents and few reciprocators ; 2) when there are enough reciprocators, altruists
become viable and can invade the population. For large populations, these three types are consequently a minimal set for emergence of cooperation starting from an all-selfish population. An example of such emergence is given by simulation of figure 6.

\section{Limits of the model}
The modeling activity necessarily simplifies real phenomena.  Modelers have to make choices, and the resulting models are not necessarily commensurable.  However, we will highlight some of the limits of our model  that future works could eventually address, and contrast, in the next section, our results to the existing literature.

First of all, our model assumes that population size remains constant throughout the co-evolutionary process. This is clearly a simplification, and the role of a changing population size remains an open problem.

The model also assumes that agents know the proportions of other agents types in the population and take their decision accordingly. Although this is a standard assumption, it would have been more realistic to assume that agents learn these proportions during their interactions, updating their belief with a bayesian rule for example. It is not clear at first glance whether these refinements would be favorable or detrimental to cooperation.

An other issue is the choice of the initial set of strategies. This choice is  itself influenced by the mode of strategy update (\textit{e.g.} replicator dynamics \textit{vs.} imitation of the most successful  agent).   We have considered  here a set based on the two most common rules, All-D and All-C, and we can question the stability of our results toward a modifications of the strategy space and interactions procedures. There have been a huge variety of strategies proposed so far in PD games and there is few chances that we could get general results about stability of cooperation independently of a precise definition of the strategy space. For example, \cite{grim1997undecidability}  demonstrated t that the outcome of a spatial prisoner's dilemma was undecidable: we know that there is no general procedure to decide whether a given strategy will invade a given ecology. The most important conclusion of theorems 1-3 is thus: in presence of altruist and selfish agents, a minimal set for emergence of cooperation has at least three strategies.

Nevertheless we can provide some intuitions about the stability and robustness our result concerning emergence of cooperation against some alternative choices. First, we can wonder whether cooperation would still be sustainable in case of a simultaneous PD. Simultaneous PD would require completely different settings since, for example, the implementation of a reciprocating strategy would imply that agents have at least a one step memory for all interaction outcomes (in our model agents only keep in memory successful first move interactions and use the sequential aspect of the interaction to eventually reciprocate when playing as second mover). But as highlighted before, the success of cooperation is due to the high popularity of altruists. This means that the important thing is that altruists can be recognized by other altruists and reciprocators in first move interactions so that offspring of these two populations make friends quicker than other types. This is still compatible with simultaneous PD, since it is possible to detect altruistic behaviors in simultaneous PD game. Nevertheless, further studies are needed to confirm this intuition.

Second, we can wonder whether cooperation would still be sustainable if we were to introduce errors in  implementation of actions. It is well known that this kind of errors is detrimental to cooperation.  However, we conjecture that while errors will certainly decrease the rate of cooperation in a population, cooperation will still be sustainable for a large domains of the parameters space when the level of errors is no too high. In our model, agents are bound to meet several cooperators because of errors in the partner selection process. Consequently their address book quickly has few names. Errors on  implementation of actions will certainly shorten this list of relationships, by discarding cooperators that would have played $D$ by error. However, if not too frequent, they should not qualitatively change the distribution of non-empty address books and preserve patterns of interaction. However, the interplay between errors on actions, errors level on partners selection and cooperation is highly none trivial and is worth being further investigated.

Last, we can question the choice of a replicators dynamics for the evolution of agent's types.  The replicator dynamics is a standard in the modeling of social dilemma. Although we can question its relevance in the modeling of cooperation in the case of human species (see for example \cite{chav:metamime}), it is a good framework to study the role of emerging social structures in situations of social dilemma where cooperation is a dominated strategy.

\section{Comparison with related work}

We have presented here a model of \textit{emergence of cooperation} in evolutionary games that highlights the role of \textit{network formation} and effects of \textit{network structure}. In line with empirical data, it proposes a mechanism that explains the \textit{persistence of heterogeneous types} (heterogeneity in rule for changing behavior), and in particular the \textit{sustainability of altruism} (presence of unconditional cooperators) even in case of \textit{strong social dilemma}. We used a \textit{hybrid approach} with both analytical and computational insights. 	
We will now discuss the contributions of our model in the light of these different points.

Emergence of cooperation occurs when a population of cooperators invades a population of defectors. This is  an important issue, both from the methodological point of view - models of cooperations should not assume \textit{a priori} that cooperation is already there -, and from an empirical point of view - lots of natural ecosystems are thought to have evolved from non cooperative states (unicellular organisms to multicellular, individuals to societies, etc.). Many explanations of cooperation rely on a cliquishness effect, which holds only once the proportion of cooperators is sufficiently large. For example, \cite{Luthi2008Cooperation}, \cite{Lozano2008Mesoscopic}, \cite{Eguiluz2005Cooperation}, \cite{Eguiluz2005Cooperation}, \cite{Fort2008Evolutionary} and \cite{Santos2006Cooperation} consider evolutions where the initial proportion of cooperators in the population is at least $50\%$. Thus, these models, without further results, are not appropriate to study the origin of cooperation in natural systems. On the contrary, our model proposes a clear scenario for emergence of cooperation, which moreover predicts different steps for the resilience of cooperative states: a slow neutral evolution with the introduction of reciprocators followed, by a faster evolution with the introduction of altruists.

Another nice property of our model is that it explains cooperation even in cases of \textit{strong social dilemma} where the temptation to defect is high ($p>0$), \textit{i.e.}  for parametrization of the social dilemma game where both the ratio between (D,C) \textit{vs.} (C,C) payoffs and (D,D) \textit{vs. }(C,D) are high. For example, in \cite{Lozano2008Mesoscopic} or \cite{Poncela2007Robustness}, the dilemma is weak since the difference between (D,D) v.s (C,D) payoffs is an $\epsilon\ll0$. In other models (\textit{e.g.} the seminal model of \cite{nowa:evol}), cooperation collapses as soon the ratio between (D,C) payoffs and (C,C) payoffs is too high.

Network formation and effect of network structure is a quite recent concern in social dilemma modeling. We can distinguish models that consider static networks with different topologies (\cite{Luthi2008Cooperation}, \cite{Lozano2008Mesoscopic}, \cite{Poncela2007Robustness}, \cite{Fort2008Evolutionary}, \cite{masu:spat}, \cite{duran2005evolutionary}, \cite{watt:small}) from models that take into account networks dynamics.

Among these later,  one of the most common  process considered for network formation is the choice and refusal mechanism (\cite{ashl:pref}, \cite{Devos1997Reciprocal}, \cite{bata:evol}, \cite{hrus:frie}, \cite{Eguiluz2005Cooperation}, \cite{Santos2006Cooperation}). Players select their partners in function of the productivity of the interaction and can refuse to interact with a player if the expected associated payoffs fall below a given threshold. In those models, the main conclusions is that selection of interactions with choice and refusal enables cooperators to interact preferentially with other cooperators, which provides the network with the appropriate level of cliquishness to sustain cooperation. This contrasts with our explanation in terms of popularity of cooperating agents.

This explanation is related to the number of interactions per agents rather than to the possibility to form cliquish communities. A similar explanation have been proposed in \cite{Pacheco2006Active}. However, there are substantial differences with our approach. \cite{Pacheco2006Active} consider a mixed population of agents of types $A$ and $B$ that can create or prune links at different rates, and interact with all the agents they are linked to at a given period. At each time step,  a link between an agent of type $i$ and an agent of type $j$ is created (resp. pruned) with probably $\alpha_{ij}$ (resp. $\beta_{ij}$). Thus, this approach can be said \textit{link centered} whereas our can be said \textit{agent centered}. The advantage of the agent centered view is that it takes into account time constraints. Think for example at sending personalized e-mail. It takes time to interact and one cannot take an arbitrary number of initiatives within a limited period of time.  This is the reason why we choose to limit to one the number of interactions for first movers.  The second important difference with \cite{Pacheco2006Active} is that the authors propose results concerning sustainability of cooperation in the domain where the timescales between the network update and the strategy update are very different ($\theta\ll1$ or $\theta\approx1$). This enables to neglect the process of link formation through time. Thus, they conclude that "the linking dynamics introduces a simple
transformation of the payoff matrix" of the prisoner's dilemma game. We saw that, at least in our case, this approximation does not hold if the condition ($\theta\ll1$ or $\theta\approx1$) is not met. In that case, it is necessary to take into account the ontogeny of agents' networks, and expected payoffs are complicated functions of $\delta$, $\gamma$ and $\theta$.

Last, contrary to all the above models, our approach introduces errors or perturbations in the choice of partners through the parameter $e$. Up to our knowledge, this dimension of the interactions between agents has not been addressed in previous models. For example,  this could bring new insights into mobility effect (high $e$ being correlated with high mobility): contrary to \cite{vainstein2007does}, we would conclude that high mobility  is detrimental to cooperation.

To conclude this section,  we would like to  finish  on a methodological point. Most approach mentioned so far present either computational studies or analytical solutions of a simplified version of the initial problem. For this study, we pay a particular attention to highlight the convergence between simulations and analytical insights.  Such a dual approach is desirable in case of models that might present internal ESS. Indeed, in case of heterogeneous populations under slow evolutionary process (which is typically what happens around an internal ESS), computational models alone are poorly informative as long as there is no way, after a finite number of iterations, to assess how close a simulation is from the attractor. With a computational approach alone, one cannot be sure that the heterogeneity observed
in the population is not an artefact due to an insufficient number of iterations. In these cases, analytical insights give supplementary evidences against the artefact hypothesis and strongly supports the understanding of the model.

\section{Conclusions}
 We took the well-known paradox of the PD to illustrate the importance of network formation in the understanding of social phenomena. We proposed a models where heterogeneous agents modify their neighborhoods according to their types, while the global network configuration determines the fitness of the different types. This co-evolution between agents' types and network patterns leads to characteristic network structures that depart from the traditional cliquish structures that account for heterogeneous cooperative equilibria. In our case, it is an increase in activity of popular (cooperative) agents that explains the emergence and sustainability of cooperation.

\section*{Acknowledgments}
These researches where supported by the Ecole Polytechnique, Paris, France, the CNRS and the Paris Ile-de-France Complex Systems Institute.
\appendix
\section{Mathematical Appendix}
\subsection{Proof of theorem \ref{theo1}}
\begin{pf}{\ref{theo1}}

We will note  $\triangle = \{ (\gamma,\delta)|\delta+\gamma \leq 1 ; \gamma \geq 1 ; \delta \geq 1 \}$ and $n_i$  the number of agents of type $i$. To compute the
mean payoffs of each type we need first to compute the probability
for an agent to be connected.

\bigskip

\begin{lem}[Theorem \ref{theo1}]{\label{lemma1}}

 Let $P_1$  be a sub-population of agents that do relationships (out-coming links) only with agents from a sub-population $P_2$ in proportion $\lambda$. The probability that a random agent $i \in P$ has already a relationship in $P_2$ is $P_{connect}(\lambda)=1-\frac{1-\theta}{1-(1-\lambda)\theta}$

\begin{pf}{}
Let $i$ be an agent of age $t$ in $P_1$. The probability it hasn't met
any agent from $P_2$ during its life is $(1-\lambda)^t$ . Consequently, the probability that agent $i$ of age $t$ hasn't any relationship in $P_2$ is $1-(1-\lambda)^t$. (We neglect here cases where the sole relationship of an agent dies before it meet an other agent from $P_2$. This is a good approximation as soon as $e.\theta. \lambda$ is not too small).

The distribution of ages in the population has the generatrice
function $f(z)=(1-\theta)\sum_{t>0}\theta ^t . z^t$. Then the probability for an agent to be connected has
the generatrice function $\Phi(z)(1-\frac{1-\theta}{1-(1-lambda)\theta}z+1-\frac{1-\theta}{1-(1-lambda)\theta})$. We deduce the mean connectedness with agents of the considered types(s):
$$P_{connect}(\lambda)=1-\frac{1-\theta}{1-(1-\lambda)\theta})$$
and its variance is :
$$Var_{connect}(\lambda)=(1-\frac{1-\theta}{1-(1-\lambda)\theta})\frac{1-\theta}{1-(1-\lambda)\theta}$$
\end{pf}
\end{lem}

    In the area $\delta>\frac{p}{r}+\frac{(1-p-r)\gamma}{r}$, the domain concerned by emergence of cooperation, only altruists cooperate first move. Consequently, they are the only one to be able to get new relationships with altruist and reciprocators. Their probability to have a non empty address book is then  $P_{connect}(\gamma+\delta)$. Selfish agents and reciprocators, on the contrary, can only do relationships with altruists. Their probability to be connected is then  $P_{connect}(\gamma)$. We can now compute the mean payoffs for each type. Payoffs terms are composed of two independent terms: the expected payoffs as first mover and the expected payoffs as second mover. For each of them we have to treat separately cases where the agent is connected, cases where it is not, and exploration moves. We thus get:
\footnotesize{
\begin{eqnarray*}
    \pi_{selfish}=&P_{connect}(\gamma)[(1-e)+e.p(1-\gamma)+e.\gamma]  ...  & \# 1 \\
     &+ (1-P_{connect}t(\gamma))[p.(1-\gamma)+\gamma] ...    &\#2  \\
     &+ e.[P_{connect}(\gamma).(1-\gamma).p+P_{connect}(\gamma+\delta).\gamma] ... &\#3  \\
     &+ (1-P_{connect}(\gamma)).(1-\gamma).p+(1-P_{connect}(\gamma+\delta)).\gamma  &\#4
\end{eqnarray*}
}
\textbf{With :}
\begin{itemize}
  \item   \#1 : $i$ is connected, 1rst move interactions
  \item  \#2 : $i$ is not connected, 1rst move interactions at random
  \item  \#3 : 2nd move interactions with connected agents
  \item  \#4 : 2nd  move interactions with not connected agents
\end{itemize}

After simplifications we get:
$$\pi_{s}=(1-e)(\frac{2p(1-\gamma+\theta(\gamma-1))+\gamma}{1+\theta(\gamma-1)}+\frac{(1-\theta)\gamma}{1-(1-(\delta+\gamma))\theta})+2e(p(1-\gamma)+\gamma)$$

Similarly we have :
\footnotesize{
\begin{eqnarray*}
\pi_{reciprocator}=&P_{connect}(\gamma)((1-e).r+e.(\gamma+p(1-\gamma)))+ ... &\# 1 \\
&+ (1-P_{connect}(\gamma))(p.(1-\gamma)+\gamma)+ ... &\#2 \\
&+e.(P_{connect}(\gamma).(1-\gamma).p+P_{connect}(\gamma+\delta).[(1-e).r.\gamma/(\gamma+\delta)+e.\gamma.r])+... &\# 3 \\
&+ (1-P_{connect}(\gamma)).(1-\gamma).p+ (1-P_{connect}(\gamma+\delta)).\gamma.r  &\#4 \\
\end{eqnarray*}
}
 After simplification we get :

$$\pi_{r} =(1-e)(\frac{\theta\gamma(r+2p-1)+\gamma+2p(1-\gamma-\theta)}{1+\theta(\gamma-1)}+\frac{r\gamma}{1-(1-(\delta+\gamma))\theta})+e(2p+\gamma(1-2p+r))\\$$

And finally :
\footnotesize{
\begin{eqnarray*}
\pi_{altruists}=&P_{connect}(\gamma+d)[(1-e).r+e.r.(\gamma+d)]+ ... &\# 1 \\
 &(1- P_{connect}(\gamma+d)).(d+\gamma).r+ ... &\# 2 \\
 &P_{connect}(\gamma).(1-e).r.d/g +Pc{connect}(\gamma+d).[(1-e).r.\gamma/(\gamma+d)+e.\gamma.r]+ ... &\# 3 \\
 &(1-P_{connect}(\gamma+d)).\gamma.r+ ... &\# 4 \\
 \end{eqnarray*}
 After simplification we get :
 $$     \pi_{a} = (1-e)(\frac{r\theta\delta}{1+\theta(\gamma-1)}+ \frac{r(2\gamma+\delta)}{1-(1-(\delta+\gamma))\theta})+er(\delta+2\gamma)$$
}

\end{pf}
\subsection{Theoretical activity}

 To compute the theoretical mean activity, we follow the same guidelines as for the expected payoffs, the
difference being that we have to take into account interactions with null payoffs.

The mean activity of agents can be split into two independents terms:
the mean activity as first mover and the mean activity as second
mover. For each of these terms, we have to treat separately cases where
the agent is connected, cases where it is not, and exploration
moves. We thus get:
\footnotesize{
\begin{eqnarray*}
Act_{s}=& 1+ ... &\# 1+\# 2 \\
&    e.[Pconnect(\gamma).(1-\gamma)+ Pconnect(\gamma+\delta).\gamma]+ ...&\# 3 \\
&  (1-Pconnect(\gamma)).(1-\gamma)+(1- Pconnect(\gamma+\delta)).\gamma+ ... &\# 4 \\
\end{eqnarray*}
}
\begin{itemize}
  \item   \#1 : $i$ is connected, 1rst move interactions
  \item  \#2 : $i$ is not connected, 1rst move interactions at random
  \item  \#3 : 2nd move interactions with connected agents
  \item  \#4 : 2nd  move interactions with not connected agents
\end{itemize}

 Similarly for reciprocators:
 \footnotesize{
\begin{eqnarray*}
Act_{r}=& 1+ ... &\# 1+\# 2 \\
&+e.P_{connect}(\gamma).(1-\gamma)+ P_{connect}(\gamma+\delta).[(1-e).\gamma/(\gamma+\delta)+e.\gamma])+ ... &\# 3 \\
&+(1-P_{connect}(\gamma)).(1-\gamma)+(1- P_{connect}(\gamma+\delta)).\gamma+ ... &\# 4 \\
\end{eqnarray*}
}
and for altruists:
\footnotesize{
\begin{eqnarray*}
Act_{a}=& 1+ ... &\# 1+\# 2 \\
&+ P_{connect}(\gamma).[e.(1-\gamma)+(1-e). d/\gamma]+P_{connect}(\gamma+\delta).[e.\gamma+(1-e). \gamma/(\gamma+\delta)]+ ... &\# 3 \\
& +(1-P_{connect}(\gamma)).(1-\gamma)+(1- P_{connect}(\gamma+\delta)).\gamma+ ... &\# 4 \\
\end{eqnarray*}
}

The relative activity of altruists (resp. reciprocators) is thus:  $RA_a=\frac{Act_{a}}{Act_{s}}$ (resp. $RA_r=\frac{Act_{r}}{Act_{s}}$).
\newpage
\section{Computational Appendix}

Lets $P(t)$ be the population at time t. We give here the algorithm used for the simulations.\\

\textbf{Parameters and notations}
The parameters are:
\begin{itemize}
  \item $e\in [0,1]$: Exploration parameter
  \item $(p,r) \in [0,\frac{1}{2}]^2$: parameter of the social dilemma
  \item $\theta \in [0,1]$: parameter defining the mean lifetime expectancy.
\end{itemize}

\bigskip

\textbf{Initial conditions }:
$N$ agents with empty address books from the three types: altruist, reciprocator and selfish, in proportion  $\gamma$ ,$\delta$   and $(1-\delta-\gamma)$ .

\bigskip

\textbf{At each period of time t :}\\
\textit{Set all payoffs to 0}\\
\textit{Interaction (parallel updating):}\\
\begin{enumerate}
\item For each agent in $P(t)$, a binomial random variable of mean $(1-e)$ is drawn to determine if its partner will  be chosen in its address book or at random in the population. In all cases, if its address book is empty, the partner is chosen at random.
  \item In case of interaction as first mover with an unknown partner, the partner is chosen at random and the action is determined according to table 2.
  \item In case of interaction as first mover in address book, the action is determined according to table 3:
  \begin{itemize}
    \item Altruists and reciprocators choose a relationship at random and play $C$.
    \item Selfish look for an altruist and defect. If they only know reciprocators, they choose one at random and cooperate. In all other cases, their interact with an agent taken at random in whole population
  \end{itemize}	
  \item Each agent plays as second mover according to its type (\emph{cf.} table 3).
  \item Each first mover adds the name of its second mover partner in its address book if the partner has cooperated. Selfish have special categories: \textit{altruist} (obtained after a $(D,C)$ results) - \textit{reciprocator or altruist} (obtained after a $(C,C)$ result) - \textit{reciprocator } (obtained after a $(D,D)$ result after interaction with an agent of the category \textit{reciprocator or altruist}.
\end{enumerate}

\textbf{Selection :}
\begin{enumerate}
  \item After interactions took place, the fitness of each agent is given by the sum of all dyadic interactions payoffs for this period.
  \item Each agent dies with a probability $(1-\theta)$ .
  \item Died agents are replaced with new agents with empty address books. Died agents are removed from the address books of their friends. This leads to the population $P(t+1)$.
  \item Types of new agents are determined by a replicator dynamics indexed on the total payoffs of agents from $P(t)$: for each new born agent,  an agent is selected in $P(t)$ with a probability proportional to its fitness. The new agent inherits the type of this particular agent.
\end{enumerate}


\bibliographystyle{jtb}
\bibliography{davidchavalarias}

\end{document}